\documentstyle[12pt]{article}
\def\btabl{\begin{table}}   \def\etabl{\end{table}}
\def\bea{\begin{eqnarray}}   \def\eea{\end{eqnarray}}
\def\bnn{\begin{eqnarray*}}   \def\enn{\end{eqnarray*}}
\def\beq{\begin{equation}}   \def\eeq{\end{equation}}  
\def\btabu{\begin{tabular}}   \def\etabu{\end{tabular}}
\def\bec{\begin{displaymath}} \def\eec{\end{displaymath}}
\def\nn{\nonumber}
\def\eqref#1{(\ref{#1})}
\begin{document}

\begin{titlepage}
\begin{flushright} FANSE-96/10\\ hep-ph/9610290
\end{flushright}

\begin{centering}

{\large \bf Do the Profile Function singularities 
explain the high energy reflection of fermions in a phase transition?}
\bigskip

{\large Jos\'e Rodr\'{\i}guez-Quintero\footnote{E-mail: jquinter@cica.es}}

\bigskip

\noindent
Departamento de F\'{\i}sica At\'omica, Molecular y Nuclear, Universidad de 
Sevilla, Spain\footnote{Work partially supported by Spanish CICYT, project 
PB 92-0663}\\
\bigskip

\end{centering}

\begin{abstract}

We investigate the scope of a previous result concerning the behaviour
of fermions hitting a general wall caused by a first-order phase transition. 
The wall profile function was considered to be analytic in the real axis. 
The previous result is valid for analytic functions in the whole complex 
plane except in certain isolated singularities located out of the real axis. 
A non-analytic profile function in the real axis is studied in order to show 
the validity of the result for any profile which can be put as a certain limit 
of a function which verifies the latter. A new understanding
of the high energy behaviour of the quantum reflection caused by a sharp
profile, as the step, arises from that study.

\bigskip
\bigskip

\noindent PACS. \ :

\begin{itemize}

\item{11.10Q-}Field theory, relativistic wave equations.
\item{11.80F-}Relativistic scattering theory, approximations.
\item{11.30Q-}Symmetry and conservation laws, spontaneous symmetry breaking.

\end{itemize}
\end{abstract}
\medskip

\end{titlepage}

The scattering of high energy fermions by a wall separating two phases of
different symmetry properties has received much attention recently . The main
physical motivation of these works is the idea\cite{kuzmin} that the baryon
asymmetry of the Universe might have been produced if the cosmological 
electroweak phase transition has been of first order. The transition is
described in terms of bubbles of a spontaneously broken symmetry vacuum
expanding in a preexisting symmetric one. In this scenario
the point to elucidate is whether there exists a CP-asymmetry that produces a
different reflection and transmission probability for quarks and antiquarks in
order to explain, via the standard model baryon number anomaly\cite{thooft},
the correct baryon asymmetry of the Universe. In order to simplify the 
treatment, an useful assumption is to break down the process into two steps, one
describing the production of CP asymmetry when the quarks/antiquarks are
reflected on the wall, the second describing the transport and the eventual 
transformation of the CP asymmetry into a baryon asymmetry. The first of two
steps justifies the effort concentrated in the study of the scattering of 
fermions in the presence of first order phase 
transition\cite{Aya94},\cite{Fun94},\cite{Rod96}.
The structure of the wall depends on the Higgs field effective potential that
takes into account the effects of the surrounding plasma trough the 
temperature of a certain thermal bath. The wall profile obtained by solving
the equation of motion with this effective potential is rather complex and
depends on too many coupling constants\cite{Nel92}, thus the study of the
general wall profile problem\cite{Fun94}\cite{Rod96} is justified not only 
from a purely formal interest. We have shown in a previous work\cite{Rod96} the
connection between the complex plane poles of the wall profile function and the
behaviour of fermions hitting the wall in the high energy limit. Nevertheless, 
by performing an extreme simplification which allows to compute the Feynman 
fermion propagator in an exact way \cite{Gav94}, the wall profile can be 
approximated
by a step function, i.e. a sudden jump from one phase to the other, that is 
called the thin wall approximation\cite{Gav94}, \cite{shapo}. 
In this case, the reflection coefficient in the high energy limit 
is\footnote{This result can be immediately
obtained from the reflection coefficient given in \cite{Gav94}, for instance. 
High energy limit means $E \gg m_0$},

\beq
R(E)={m_0\over 2E} \ ,
\label{1}
\eeq

\noindent being the mass $m_0$ the high of the wall. Notice the power law
dependence on $E$ of the reflection coefficient in eq. (\ref{1}). 
As we will see, the results given in ref. \cite{Rod96} cannot be directly 
applied to profile functions which are not analytic in the real axis, as the
step profile function. Nevertheless, in this note we show the high energy
behaviour of a new kind of profile functions, which can be considered in a
certain limit as analytic extensions in the complex plane of the step function,
by following \cite{Rod96}. Furthermore, we will investigate the appropriate
limit in order to understand the result (\ref{1}) through 
the analytic properties of the step function extensions.

Next, we outline the derivation of the high energy asymptotic expression for 
the reflection coefficient given in ref. \cite{Rod96}, where the detailed
calculation can be found.

By formulating the problem in the rest frame of a wall normal to the $z$-axis,
characterized by a general non-CP-violating wall profile\footnote{CP-violating 
wall profiles are studied in \cite{Rod296}}, and working in the chiral 
basis\cite{Nel92}, we can factor the Dirac equation into $2 \times 2$ identical 
blocks. Thus, the problem is reduced to solve the following equation,

\beq
\left ( i \partial_z + Q(z) \right ) \psi_{I/II} = 0 \ ,
\label{2}
\eeq

\noindent with 

\beq
{\em Q}(z) = \left( \begin{array}{cc} E & -m(z) \\ m(z) & -E \end{array}
\right) \;\;\;\; \mbox{and} \;\;\;\; \Psi = {\psi_I \choose \psi_{II}} e^{-iEt}
\ ; \label{3}
\eeq

\noindent where $\Psi$ is the time-independent Dirac equation solution in the
chiral basis.
We express the solution of (\ref{3}) as follows

\beq
\psi_{I/II}(z) = \Omega(z,z_0) \left( 
\begin{array}{c} \psi_1(z_0) \\ \psi_2(z_0) \end{array} \right) \ ,
\label{4}
\eeq

\noindent with

\beq
\Omega(z,z_0)={\em P}e^{i\int^z_{z_0}d\tau {\em Q}(\tau)} \ ;   
\label{5}
\eeq

\noindent where ${\em P}$ indicates a path ordered product and $\tau$ is the
position variable along the z-axis. We consider $m(\tau)=m_0f(\tau)$, where
$f(\tau)$ is the profile wall function. The asymptotic conditions
$f(+\infty)=1$ and $f(-\infty)=0$ are required, $f(\tau)-\theta(\tau)$ 
decreasing exponentially when $\tau \to \pm\infty$. Thus, after some tedious
calculation we obtain from equations (\ref{3}) and (\ref{5})

\bea
\Omega (z, z_0) = \left( 1+ \left( {m_0 \over E}\right)\sigma_2 \int_{z_0}^z 
d\tau \ p(\tau ) \ f(\tau ) \ e^{-2i\sigma_3\int_{\tau}^z d\xi p(\xi)} \right.
\nn \\
+ \left. O\left[\left({1\over E}\right)^2\right] \right) 
e^{i\sigma_3 \int_{z_0}^z d \tau \ p(\tau )} \ ; \label{6} 
\eea

\noindent where $p(\tau)=+\left( E^2-\left[m_0 f(\tau)\right]^2\right)^{1/2}$.

Following Nelson {\it et al}\cite{Nel92}, we can obtain the reflection 
coefficient from the result (\ref{6}). Thus, by assuming in general that

\beq
f(\tau)=F\left( {\tau\over \sigma} \right) \ ,
\label{7}
\eeq

\noindent where the parameter $\sigma$ gives the wall thickness, and after much 
calculation\cite{Rod96}, we finally obtain

\beq
R(E) = 2 \pi \sigma m_0 \sum_{j=1}^N e^{-2E\sigma y_j} \
e^{2iE\sigma x_j} \sum_{n=1}^{\nu_j} b_{-n}^j 
{(2iE \sigma )^{n-1} \over (n - 1) !} \ ; 
\label{8}
\eeq

\noindent where $z_j = x_j + i y_j$ and $b_{-n}^j$, for $j=1, ..., N$, are all 
the poles of $F(z)$ with positive imaginary part and the n-power coefficient 
of the Laurent expansion for the function in each pole, respectively. The order
of the pole $z_j$ is $\nu_j$.

The result (\ref{8}) is valid in general for profile functions which are 
analytic in the real axis, i.e. analytic functions in all the complex plane
except in certain isolated singular points with non-zero imaginary part. 
It is positively checked in ref. \cite{Rod96} by using the particular 
Kink-type wall profile\cite{Aya94}\cite{Fun94}, 
$f(\tau)={1\over 2}\left( 1 + \tanh(\tau/\sigma) \right)$. We consider now 
the more general kind of functions 

\beq
f(\tau)= {1\over 1 + \exp\left(-{\tau^{2n+1}\over \sigma}\right)} \ .
\label{9}
\eeq

\noindent It is obvious that the functional behaviour of the Kink-type wall is 
a particular case of the latter, given by taking $n=0$. Moreover, any of these
functions gives the step function if the limit $\sigma \to 0$ is taken.
In order to apply the
result (\ref{8}) to the functions (\ref{9}), we must know the singularities
of these functions and their distribution in the complex plane. Taking into 
account eq. (\ref{7}), we obtain for the poles of $F(z)$

\beq
z_{jk} = \pi^{{1\over m}}(1+2j)^{{1\over m}}
e^{\pm i{\pi\over m}({1\over 2}+2k)} \ ;
\label{10}
\eeq

\noindent with $j=0,1,2, \ ... \ \infty$ and $k=0,1, \ ... \ m-1$. Where $m=2n+1$ 
is the power in the exponential argument in (\ref{9}). 
The order of all the poles is $1$ and the residous, $b_{-1}^{jk}$, 
can be written as 

\beq
b_{-1}^{jk} = {1\over m}\left[\pi\left(1+2j\right)\right]^{1-m\over m}
e^{\pm i{1-m\over m}\pi({1\over 2}+2k)} \ .
\label{11}
\eeq

It is easy to see that these poles are distributed in the complex plane along 
$m$ radial axes, on circumferences with radius given by 
$\left[\pi\left(1+2j\right)\right]^{1/m}$, where $j=0,1,...\infty$. In fact,
the distribution of the poles is symmetric with regard to the real axis. The
schematic location of the poles in the complex plane for $m=1,3$ is shown in
fig. 1 . It can be found that the two poles with lower positive imaginary part 
are

\beq
z_{0,0}=\pi^{{1\over m}}e^{i{\pi\over 2m}}
\;\;\;\; \mbox{and} \;\;\;\;
z_{0,{m-1\over 2}}=\pi^{{1\over m}}e^{i\pi(1-{1\over 2m})} \ ,
\label{12}
\eeq

\noindent and by applying the result (\ref{8}) in the range of energies 
$E\sigma \gg 1$, we obtain

\beq
R(E)={2(2-\delta_{m,1})\over m}\pi^{{1\over m}}\sigma m_0
\cos\left[2\pi^{{1\over m}}\cos\left({\pi\over 2m}\right)\sigma E +
{1-m\over m}{\pi\over 2}\right] \ 
e^{-2\pi^{1\over m}\sin({\pi\over 2m})\sigma E} \ ;
\label{14}
\eeq

\noindent where all the decreasing exponential terms except the two with 
lowest arguments has been neglected. The factor $2-\delta_{m,1}$ arises because
the two poles considered in eq. (\ref{12}) are the same for $m=1$. 
The exponential law of eq. (\ref{14}) is the expected behaviour for the 
reflection coefficient provided that the result (\ref{8}) can be applied, if
the energy is large enough. Therefore, the reflection coefficient for the step 
wall profile which presents a power law asymptotic behaviour, as we mentioned 
above, does not seem to be coherent with the former. Nevertheless, we must take
into account that the result (\ref{8}) relates the high energy behaviour of 
the fermions hitting a wall to the analytic properties of the profile function 
in the complex plane. Thus, we must appropriately extend the step profile in 
the complex plane to investigate if the obtained poles allow to 
explain the behaviour given by (\ref{1}). 

It can be easily seen that the limit of the functions (\ref{9}), for all $n$, 
when the parameter $\sigma$ goes to zero is the step function. Therefore, in 
that limit, any of those can be considered as an analytic extension in the 
complex plane for the step function. By considering that the function $F(\tau)$
is the same than $f(\tau)$, but scaled by the parameter $\sigma$, we have the 
same pattern for the distribution of poles of $f(\tau)$ than for that of
$F(\tau)$, but with the distance between two consecutive poles depending on
$\sigma$. In other words, the poles will be as close to each other along the
same radial axes as $\sigma$ small. The result (\ref{8}) cannot be strictly
applied for $\sigma=0$, but we know that if the limit when $\sigma \to 0$ 
of the high energy reflection coefficient exists, it must be the same 
obtained by using (\ref{8}), which is valid for any non-zero $\sigma$. 
The crucial point in this approach is that when $\sigma$ is getting smaller 
and smaller we cannot consider a large enough energy, $E$, to neglect the 
contribution other than the closest poles to the real axis. In this way, the 
contributions due to each pole have to be summed before to take the limit and 
there is no reason to expect an exponential final result.

Indeed, we have

\beq
R(E) = 2 \pi \sigma m_0 \sum_{k=0}^{m-1}\sum_{j=0}^N e^{-2E\sigma y_j} \
e^{2iE\sigma x_j}  b_{-1}^{jk}\ , 
\label{15}
\eeq

\noindent where $b_{-1}^{jk}$ is defined in eq. (\ref{11}) and 
$y_{jk}$, $x_{jk}$ are the real and pure imaginary components of $z_{jk}$, 
given by eq. (\ref{10}). By taking $\sigma \to 0$ in eq. (\ref{14}), the
sum for $j$ from $0$ to $+\infty$ can be considered as an integral and after
some appropriate transformations, we obtain

\beq
R(E)={m_0\over 2E} \ \sum_{k=0}^{m-1} \ e^{i\beta_k}
\int_0^{\infty}dt \ e^{-t\sin\alpha_k}e^{it\cos\alpha_k} \ ;
\label{16}
\eeq

\noindent where

\bea
&&\beta_k = \pm {\pi\over m}(1-m)({1\over 2}+2k) \ , \nn \\
&&\alpha_k = \pm {\pi\over m}({1\over 2}+2k) \ .
\label{17}
\eea

\noindent As we only sum for the poles with positive imaginary part, the signs
$+,-$ in eq. (\ref{17}) are for $k \leq {m-1\over 2}$ and $k>{m-1\over 2}$,
respectively. By solving eq. (\ref{16}), we obtain

\bea
R(E)={m_0\over 2E} \ \sum_{k=0}^{m-1} e^{i(\beta_k+{\pi\over 2}-\alpha_k)}
={m_0\over 2E} \left\{ {m+1\over 2} - {m-1\over 2} \right\} \nn \\
={m_0\over 2E} \ ;
\label{18}
\eea

\noindent which agrees with eq. (\ref{1}). In consequence, the power law
behaviour which is characteristic for the step function, as well as for 
any non-continuous function, and therefore non-analytic, in the real axis,
can be well understood through the complex analytic properties of these 
functions. The high energy reflection behaviour is caused by the sum of 
the contributions due to each singularity of the profile function. For a
large enough energy the contributions of one or several of these singularities
can be considered as leading, provided that the singular points are isolated. 
We only must take into account that the real non-continuous point is
produced when the poles of the function are distributed in an infinitely
compact way along lines in the complex plane which contains this point to 
understand the power law for the step. Thus, the smoother decreasing on the
energy of the reflection coefficient for the sharp profiles can be explained
because no singularity contribution can be isolated in order to lead the total
result.

\bigskip

{\bf Aknowledgements}

\medskip

I thank O. P\`{e}ne and M. Lozano for essential comments and very helpful 
discussions. I am specially grateful to the latter for his careful reading of
the manuscript.

\newpage

\section*{Figure Captions}
\begin{itemize}

\item[Figure 1.]
Schematic distribution in the complex plane of the poles, $z_{jk}$, given by
eq. (\ref{10}), for $m=1$ (a) and for $m=3$ (b). 
The poles are located in the solid points, distributed along radial axes,
as explained in the text.

\end{itemize}

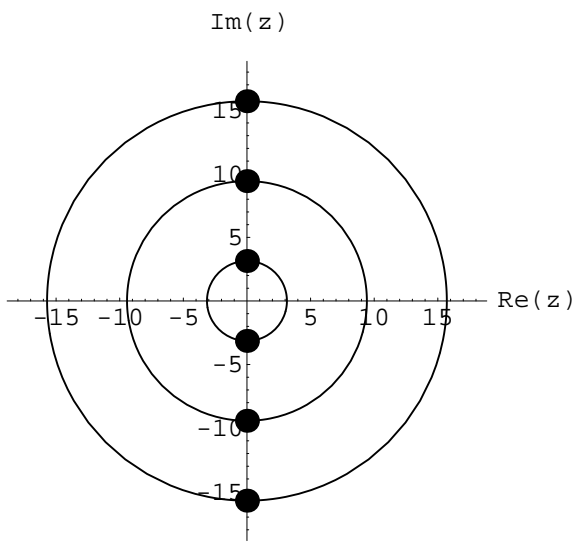
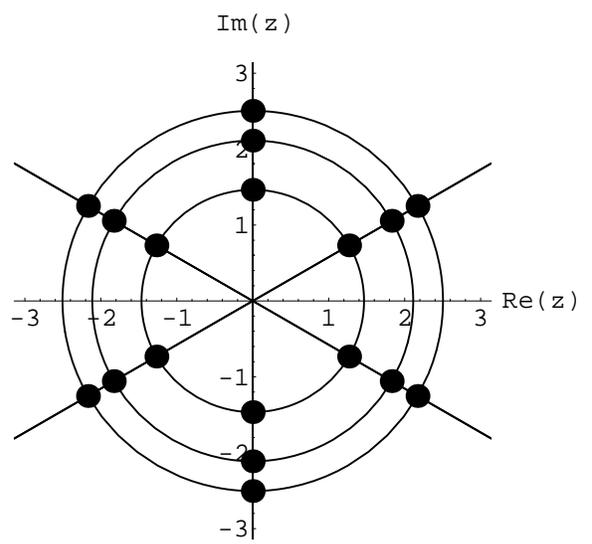
\end{document}